\documentclass{iau}

\usepackage{graphicx} \usepackage{caption} \usepackage{subcaption}

\title[STARDISK] %% give here short title %%
{The effect of gaseous accretion disk on dynamics of the stellar cluster in AGN}

\author[Bekdaulet Shukirgaliyev]   %% give here short author list %%
{Bekdaulet Shukirgaliyev$^1$
 }

\affiliation{$^1$Fesenkov Astrophysical Institute \\ Observatory 23, 050020 Almaty, Kazakhstan \\ email: {\tt bekdaulet@aphi.kz}}

\pubyear{2014}
\volume{312}  %% insert here IAU Symposium No.
\pagerange{xxx--xxx}
\jname{Star clusters and black holes in galaxies across cosmic time}
\editors{A.C. Editor, B.D. Editor \& C.E. Editor, eds.}
\begin{document}

\maketitle

\begin{abstract}
There is a super-massive black hole, a gaseous accretion disk and compact star cluster
in the center of active galactic nuclei, as known today. So the activity of AGN can be
represented as the result of interaction of these three subsystems. In this work we investigate
the dynamical interaction of a central star cluster surrounding a super-massive black hole
and a central accretion disk. The dissipative force acting on stars in the disk leads to an
asymmetry in the phase space distribution of the central star cluster due to the rotating accretion
disk. In our work we present some results of Stardisk model, where we see some changes in
density and phase space of central star cluster due to influence of rotating gaseous accretion
disk.
\end{abstract}

\section{Introduction}

	Physical nature of AGN is far from full understanding because of big distances and relatively 
	small sizes of energy-producing regions. Therefore development of AGN theory still remains one 
	of the main problems in astrophysics (\cite[Beckmann \& Shrader 2013]{BeckmannShrader13}, \cite[Wu et al. 2015]{Wu_etal2015}). 

	According to the dominant model, the phenomenon of AGN is explained by the accretion of matter onto 
	a supermassive black hole in the center of galaxies (according to modern data (\cite[Kormendy \& Ho 2013]{KormendyHo2013}), there are supermassive
	black holes with mass from a few billion to several trillions of solar masses in the center of most 
	galaxies). In the process of accretion potential and kinetic energy of substances is effectively converted
	into radiation energy, which may explain the stable and very powerful radiation from a very small 
	region, observed in AGNs. Since the angular momentum of accreting substances is retained, it forms a 
	disk, and the system becomes axisymmetric.

	However, there are typically sperically symmetric compact stellar clusters around central black hole in the 
	central region of galaxies. The previous work (\cite[Just et al. 2012]{Just_etal12}, 
	\cite[Vilkoviskij et al. 2013]{Vilk_etal13}) investigated how star-disk interactions effect
	the evolution of AGN using simpler model of accretion disk and numerical simulations. In particular, it
	was found that the dissipative star-disk interactions can significantly increase the rate of accretion of stars
	onto SMBH as stars transfer portion of the energy to gas and their orbits are stacked in the plane of 
	accretion disk, and as a result they also accreting onto the central black hole (\cite[Just et al. 2012]{Just_etal12}). However, to assess the 
	impact of the accretion disk in the orbital and phase characteristics of the stars, one needs to explore 
	a more realistic model of the disk. 
 
The investigated AGN model includes three subsystems: a compact star cluster (CSC), the accretion disk (AD), 
and the central supermassive black hole (SMBH). A star cluster is simulated by direct integration of the 
individual stars interaction with each other (direct N-body simulations) and with a gas disk and black hole. 
Gaseous accretion disk is defined phenomenologically with density distribution constant in time, and 
has a Keplerian rotation curve. Black hole is also defined phenomenologically as a Newtonian potential. If the 
particle comes inside the region whith radius less than $ R_{accr} $ (radius of accretion), then we 
consider that it is accreted onto the supermassive black hole. Once it happened, the particle disappears 
and its mass is added to the SMBH. We used modified version of phiGRAPE code (\cite[Harfst et al. 2007]{Harfst_etal07})for our simulations, which uses
 parallel computing technology of NVIDIA CUDA and MPI. The stellar component of the system is defined as a Plummer
 density profile initially. A more detailed description of the numerical model can be found in \cite[Just et al. 2012]{Just_etal12}, 
	\cite[Vilkoviskij et al. 2013]{Vilk_etal13}.

	In Kazakhstan the first computer cluster specialized for N-body simulations was created in 2008 within the 
	frame of the joint Kazakh-German "STARDISC" project which combined forces of groups in Heidelberg 
	University and Fesenkov Astrophysical Institute. This work presents results of investigations of 
	evolution and physics problems in AGN are calculated in this computer cluster as well as in computer cluster of 
	Astronomisches Rechen-Institute in Germany.

 \section{Accretion disk model}

 Let us consider the model of the accretion disk. We take as a basis a three-dimensional, axisymmetric stationary
 disk, which is characterized by differential rotation with the local angular velocity. The radial profile of 
 the surface density is specified as
\begin{equation}\label{eq1} 
\Sigma (R) = \Sigma_d\left(\frac{R}{R_d}\right)^{-\alpha},
\end{equation}
here $ \alpha = \frac{3}{4}, R^2 = x^2 + y^2, R_d $ is disk radius and $ \Sigma_d $ is surface density 
 value at $ R = R_d.$ The value $ \alpha = 3/4 $ corresponds to the outer boundary of the disk model of the 
 \cite[Novikov \& Thorne 1973]{NovikovThorne73}. Disk mass is equal to
\begin{equation}
\label{eq:2} 
M_d = 2 \pi \int_0^{R_d} \Sigma (R) RdR = {2 \pi \over 2 - \alpha} \Sigma_d R_d^2.
\end{equation}

For the numerical integration of the equations of motion the force acting on the particle should be smooth 
and continuous function, so one needs enter the exponential factor that ensures a smooth (but fast enough)
 decreasing of gas density at the edge of the disk (\cite[Just et al. 2012]{Just_etal12}).
\begin{equation} 
\label{eq:3}
\Sigma (R) = \Sigma_d\left({R \over R_d}\right)^{- \alpha} e^{-\beta_s \left({R \over R_d}\right)^s}.
\end{equation}

For the resulting expression to correspond to the equation \ref{eq:2} we selected 
$$ \beta_s = \Gamma\left(1+ {2-\alpha \over s}\right),$$ wherein $ \Gamma(x) $ - the gamma function. Let us
 take $ s = 4 $, then $beta_s = 0.70$ for $\alpha = 3/4$. In this case, if $R = R_d$ the surface density is
 equal to $\Sigma(R_d) = 0.49\Sigma_d$ (\cite[Just et al. 2012]{Just_etal12}).
 To numerically model the described model we select isothermal density profile, defined as 

\begin{equation}
\label{eq:4}
\rho_g (R,z) = \frac {\Sigma(R)} {\sqrt{2\pi}h_z} exp \left(- {z^2 \over 2h_z^2}\right).
\end{equation}

Previous works (\cite[Just et al. 2012]{Just_etal12}, \cite[Vilkoviskij et al. 2013]{Vilk_etal13}) used constant thickness accretion disk model 

\begin{equation}
\label{eq:5}
h_z = hR_d.
\end{equation}

 If the relationship \ref{eq:5} is substituted into the expression \ref{eq:4}, we obtain the expression for the density of disk with a constant height:
\begin{equation}
\label{eq:6}
\rho(R,z) = \frac{2-\alpha}{2\pi \sqrt{2\pi}}{M_d \over hR_d^3}\left({R \over R_d}\right)^{-\alpha} e^{-\beta_s\left({R \over R_d}\right)^s} e ^{-\frac{z^2}{2h^2R_d^2}}.
\end{equation}

In this work we consider a new model, which is a modification of the first model of accretion disk with the introduction of
 a linear function to increase the disk half thickness in the inner part. This modification is based on the physical 
 properties of the inner accretion disk, which is described by the approach of \cite[Shakura \& Sunyaev 1973]{ShakuraSunyaev73}
\begin{equation}
\label{eq:7}
h_z = hR_d\left({R \over R_{crit}}\right).
\end{equation}
The transition point from linearly increasing to constant thickness is in the point where the disk is vertically self-gravitating 
at $R = R_{crit}$. In our simulations $R_{crit} = 0.0257314$ in N-body dimensionless unit system (\cite[H\'enon 1971]{Henon71}).

The properties of the accretion disc are fixed by the reduced mass with analytical density distribution according to equation 
\ref{eq:6} with the values $\alpha = 3/4, s =4 $ and $h = 10^{-3}$. The disk has Keplerian rotation in the potential of SMBH 
neglecting the gravitational influence of the disk and pressure gradients within the disk (\cite[Just et al. 2012]{Just_etal12}). 

\begin{figure}[h]
  \centering
  \includegraphics[width=0.7\textwidth]{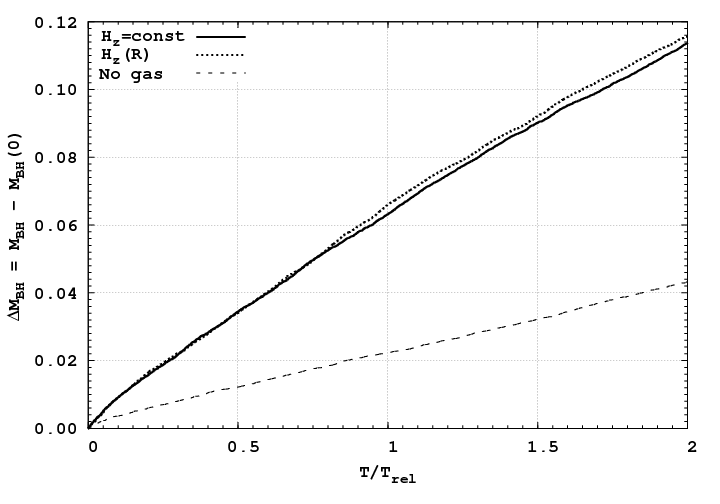}
  \caption{The black hole mass growth as a function of time for three model runs
  - the old model of disk (solid line), with the new model (dotted line) and 
  a control run without gas disk (dotted line). The mass of black hole is in dimensionless 
  N-body units, the unit of time specified in the relaxation time of the system.}
  \label{fig1}
\end{figure}

\section{Results}

To ensure that the new model of disk does not change the global dynamics of the system and thereby does
 not contradict the results we obtained earlier, we compared the black hole mass growth rate due to the 
 accretion of stars (Fig.\,\ref{fig1}).

 \begin{figure}[h]
  \centering
  \includegraphics[width=0.9\textwidth, page=3]{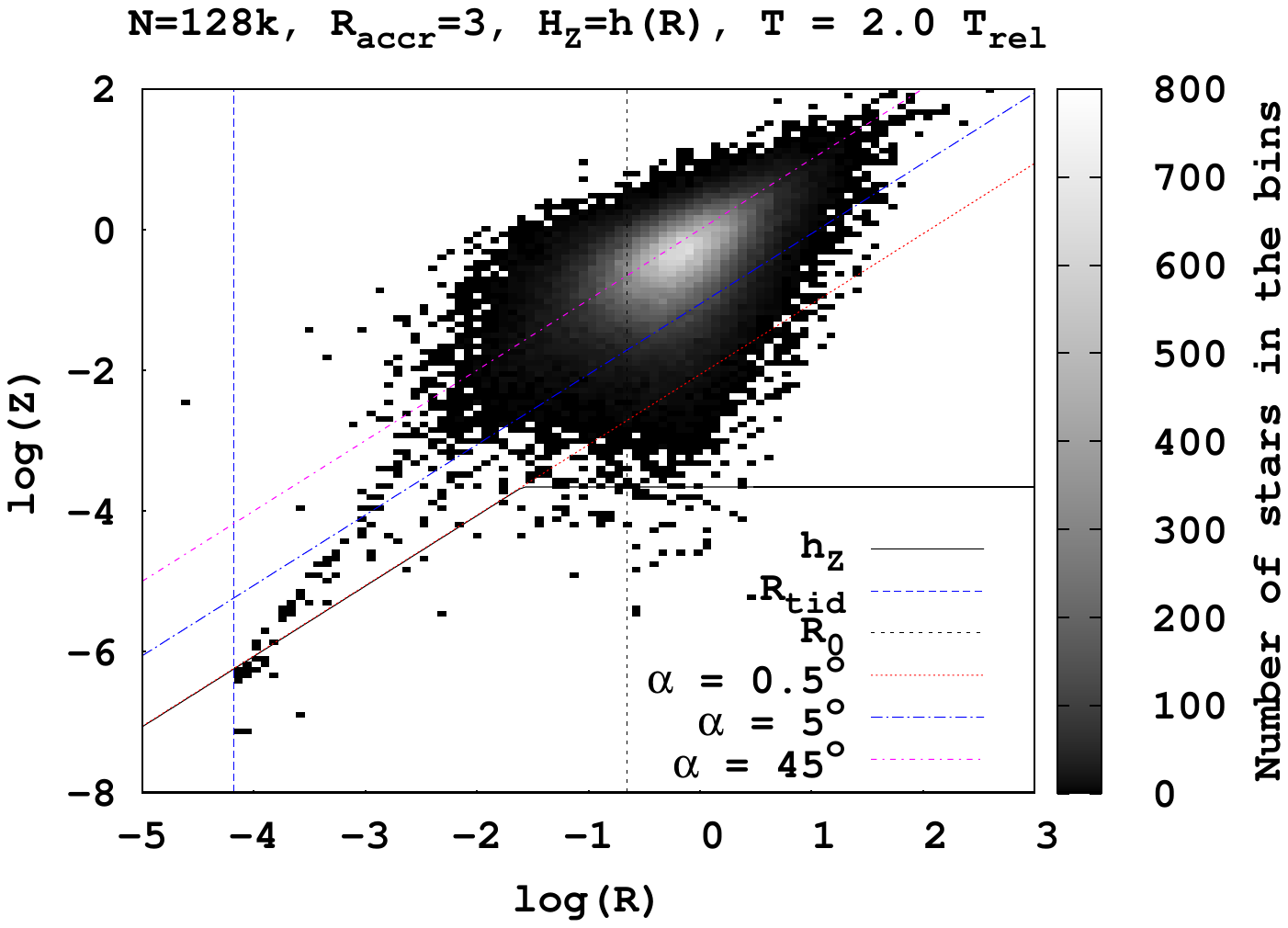}
  \caption{Stellar distribution in cylindrical coordinate system. Here stellar disk formed in the inner part of the 
  system shown as a tail to the center of the system.}
  \label{fig2}
\end{figure}

 As can be seen in Fig. 1  the evolution of the black hole during two 
 relaxation times for both accretion disk models is identical. At the same graph control run without 
 gaseous disk is shown. In this case, the black hole mass growth is due to only the capture of those stars, 
 which accidentally happened to be within the accretion radius.
We also investigate the impact of accretion disk on the dynamical characteristics of stellar cluster. In particular, 
we analysed orbital parameters of accreted particles. We found that these two disk models give different results in 
orbital parameters distribution of captured particles. Specifically, the new disk model allows stars to survive in the 
accretion disk for longer times than in old disk model. So we get much more particles accreted in near-circular orbits ($e \approx 0$) in 
the disk plane ($i < 0.5^{\circ}$). 
Although both disk models initiate the same growth rate of the central SMBH (Fig.\,\ref{fig1}), the gaseous disk with a constant 
thickness captures more counter-rotating stars, in contrast to the disk with a variable thickness. 

If we decrease the accretion radius value in the previous model of the disk, the gas density 
increases in the central part, but its thickness remains constant and it rotates around the 
center on Keplerian orbits. This leads to deceleration of many stars on orbits  counter-rotating with
the disk in the inner part of the system, including stars in nearly perpendicular orbits
 to the disk plane. In case of an improved disk model, super-dense gas occurs practically only 
 in the equatorial plane near the central black hole, and that allows many stars in the central part
 to evolve towards the direction of rotation of the disk. Stars evolved in the positive direction
 of rotation relatively to the direction of the disk rotation could increase the contribution of the positive 
 Z-component of the angular momentum in the central part of the cluster.
 
 We found that the disk model with varying thickness leads to formation of relatively stable disk from stars in the 
inner part of the system. It captures stars which are going to be accreted and allows them to live longer in the disk.
One can see formation of such disk in (Fig.\,\ref{fig1}). The central star cluster changes its structure in the inner part and forms stellar disk in the early
stage of simulation ($T \approx 0.01 T_{rel}$). This structure of star cluster retained during two relaxation times (up to the end of the simulations).

Thus, we see that the resulting phenomenological model of gas accretion disk is physically adequate 
and the studied density profile can be used in the future, including modeling of the gas disk directly
 using the methods of hydrodynamics. In the future, we plan to improve the numerical model of AGNs in order
 to allow additional processes (collisions of stars in the central part, stellar evolution,
 primordial binaries, binary black holes, etc.), and to perform one-to-one simulations, i.e. with the number of stars equal
 to that in real clusters (about one million stars).
 
I would like to thank my collaborators G. Kennedy, Y. Meiron, P. Berczik, A. Just, T. Panamarev, M. Makukov, D. Yurin, C. Omarov, E. Vilkoviski, and
 R. Spurzem. We will publish a more detailed presentation of results elsewhere.
 
I also thank International Astronomical Union for its grant which gave me this opportunity to participate in IAU Symposium No. 312 in Beijing in 2014.
 
{}
\end{document}